\documentclass[times, twoside]{zHenriquesLab-StyleBioRxiv}
\usepackage{graphicx}

\leadauthor{Feki} 

\begin{document}

\title{ConvNeXt with Histopathology-Specific Augmentations for Mitotic Figure Classification}
\shorttitle{ConvNeXt for MIDOG 2025}

\author{Hana Feki, Alice Blondel and Thomas Walter}

\affil{CBIO - Center for Computational Biology, Mines Paris PSL, France}
\maketitle

\begin{abstract}
Accurate mitotic figure classification is crucial in computational pathology, as mitotic activity informs cancer grading and patient prognosis. Distinguishing atypical mitotic figures (AMFs), which indicate higher tumor aggressiveness, from normal mitotic figures (NMFs) remains challenging due to subtle morphological differences and high intra-class variability. This task is further complicated by domain shifts—including variations in organ, tissue type, and scanner—as well as limited annotations and severe class imbalance.

To address these challenges in Track 2 of the MIDOG 2025 Challenge, we propose a solution based on the lightweight ConvNeXt architecture, trained on all available datasets (AMi-Br, AtNorM-Br, AtNorM-MD, and OMG-Octo) to maximize domain coverage. Robustness is enhanced through a histopathology-specific augmentation pipeline—including elastic and stain-specific transformations—and balanced sampling to mitigate class imbalance. A grouped 5-fold cross-validation strategy ensures reliable evaluation.

On the preliminary leaderboard, our model achieved a balanced accuracy of 0.8961, ranking among the top entries. These results highlight that broad domain exposure combined with targeted augmentation strategies is key to building accurate and generalizable mitotic figure classifiers.

\end{abstract}

\begin{keywords}
Mitotic figure classification | Domain Shift | Data augmentation | Digital pathology \\
Correspondence: \href{mailto:hana.feki@ensta.fr}{hana.feki@ensta.fr}
\end{keywords}

\section*{Introduction}  

Mitotic figure (MF) classification is a critical task in computational pathology, as mitotic activity informs cancer grading and patient prognosis. Distinguishing atypical mitotic figures (AMFs), which are associated with higher tumor aggressiveness, from normal mitotic figures (NMFs) is particularly challenging due to subtle morphological differences and high intra-class variability.
NMFs progress through well-defined stages—prophase, metaphase, anaphase, and telophase—each displaying characteristic chromosomal arrangements and cellular morphology. AMFs deviate from these canonical stages and exhibit diverse morphologies, including multipolar, lagging, or fragmented chromosomes.
Both NMFs and AMFs are further affected by high inter-domain variability: morphology can differ across organs, tumor types, staining protocols, and scanning conditions, increasing intra-class heterogeneity and complicating automated classification. Even expert pathologists face difficulties in consistently identifying AMFs, and current AI solutions often underperform on this task. Developing a classifier that performs reliably across all domains—different organs, scanners, staining protocols, and tumor types—remains an open challenge.

To mitigate the impact of domain shifts in image classification, several strategies have been explored in computational pathology. Domain-adversarial training \cite{Ganin2017} encourages models to learn features that are invariant across domains, reducing sensitivity to scanner or staining differences. Stain normalization aims to harmonize variations in tissue appearance, while color augmentation artificially increases variability during training to improve robustness. Additionally, multi-domain training exposes models to diverse datasets, allowing them to learn more generalizable features across different scanners, staining protocols, and tissue types. 

In this context, The Mitosis Domain Generalization Challenge 2025 (MIDOG25) provides a standardized benchmark for developing and evaluating mitotic figure classifiers under realistic multi-institutional variability. Track 2 of the challenge specifically focuses on mitotic figure classification, requiring models to detect and classify mitoses across diverse datasets with varying staining protocols, tumor types, and scanning conditions. 

In this work, we trained a lightweight ConvNeXt-based classifier \cite{liu2022convnet} on all available mitotic figure datasets, allowing the model to learn from diverse tumor domains and image variations. Combined with a carefully designed histopathology-specific augmentation pipeline and balanced sampling, this approach yields a robust model capable of generalizing across unseen domains, achieving state-of-the-art performance in mitotic figure classification under domain shift.

\section*{Material and Methods}

\subsection*{Datasets}

We leveraged all publicly available datasets provided by the organizers of the \textbf{MIDOG 2025 Challenge}, encompassing multiple tumor types and staining conditions to ensure broad domain coverage:
    \begin{itemize}
    \item \textbf{AMi-Br \cite{palm_histologic_2025}}: A human breast cancer dataset containing 3,720 mitotic figures, of which 22.37\% are atypical. Annotations were performed using a 3-expert majority vote, with a full agreement rate of 78.2\%. 

    \item \textbf{AtNorM-Br \cite{banerjee2025benchmarkingdeeplearningvision}}: A human breast cancer dataset derived from TCGA, containing 746 annotated mitotic figures (17.16\% atypical). Annotations were provided by a single expert. The dataset is designed with normalized staining to reduce variability across slides, supporting the development of more robust models.

    \item \textbf{MIDOG++ \cite{aubreville_comprehensive_2023}}: A multi-domain dataset containing a total of 11,939 mitotic figures, of which 14.64\% are atypical (1,748 AMFs and 10,191 NMFs), spanning 7 different tumor domains. Annotations were established through a 3-expert majority vote, with a full agreement rate of around 70\%. This dataset allows evaluation of model generalization across species and domains. 
    
    \item \textbf{OMG-Octo Atypical}: A refined version of the OMG-Octo dataset \cite{shen_omg-octo_atypical_2025}, focused on atypical mitoses. It contains 3,024 mitotic figures, including 1,378 (45.6\%) AMFs, 379 normal NMFs, 394 apoptotic cells, 399 noise, and 462 uncertain cases.  

\end{itemize}  

These datasets collectively provide a challenging multi-domain setting, reflecting the variability encountered in clinical practice.  
For training and validation, we performed a 5-fold cross-validation split across the combined datasets, ensuring balanced representation of mitotic and non-mitotic figures.

\subsection*{Model Architecture}
We employed the \textbf{ConvNeXt small} classifier \cite{liu2022convnet}, initialized with ImageNet-pretrained weights. This architecture was selected for its strong image classification performance combined with computational efficiency, featuring only 49.46M parameters. The network outputs binary logits for atypical mitotic figure classification, making it particularly well-suited for high-resolution histopathology patches, where accurate discrimination and efficient processing are both essential.

\subsection*{Data Augmentation}
To enhance robustness under domain shifts and tissue variability, we designed a comprehensive augmentation pipeline. 
Instead of applying all augmentations simultaneously to each image, the pipeline randomly selects different types of transformations at each iteration. 
This approach effectively presents the model with a wide range of plausible variations of the same image, akin to training on multiple slightly different datasets. 
By exposing the network to diverse spatial, color, and noise patterns, it learns more generalizable features and reduces overfitting.  

\begin{itemize}
    \item \textbf{Geometric transforms:} D4 rotations (all 8 combinations of 90° rotations and flips), random rotations up to 180°, and random 90° rotations (overall probability 0.9). These augmentations improve invariance to cell orientation and local tissue deformations.  

    \item \textbf{Advanced geometric transformations:} ShiftScaleRotate (shift $\pm$8\%, scale $\pm$15\%, rotate $\pm$30°, $p=0.8$), ElasticTransform ($\alpha=40$, $\sigma=4$, $\alpha_\text{affine}=8$, $p=0.7$), GridDistortion (5 steps, distort limit 0.2, $p=0.6$), and OpticalDistortion (distort limit 0.15, $p=0.5$). These simulate realistic slide-specific artifacts and tissue deformations.  

    \item \textbf{Color augmentations:} ColorJitter (brightness/contrast $\pm$0.2, saturation $\pm$0.15, hue $\pm$0.08, $p=0.8$), HueSaturationValue (hue $\pm$15, saturation $\pm$20, value $\pm$15, $p=0.8$), RandomBrightnessContrast (brightness/contrast $\pm$0.2, $p=0.8$), and CLAHE (clip limit 2.0, tile grid 4x4, $p=0.4$). These simulate staining and scanner variability.  

    \item \textbf{Channel manipulations:} RGBShift (shift $\pm$20, $p=0.6$), ChannelShuffle ($p=0.3$), and occasional grayscale conversion ($p=0.1$, overall $p=0.4$). These improve robustness to variations in color channels.  

    \item \textbf{Blur and noise:} GaussianBlur (kernel 1–5, $p=0.5$), Defocus (radius 1–4, alias blur 0.1–0.3, $p=0.4$), MotionBlur (limit 5, $p=0.3$), GaussNoise (mean=0, per channel, $p=0.4$), ISO noise (color shift 0.01–0.05, intensity 0.1–0.4, $p=0.3$), and MultiplicativeNoise (multiplier 0.95–1.05, $p=0.2$). These augmentations simulate realistic microscopy conditions, including focus variations and sensor noise.  

    \item \textbf{Final preprocessing:} $60 \times 60$ px center crop, resize to $224 \times 224$, and ImageNet normalization. Cropping focuses on mitotic regions, resizing adapts inputs to ConvNeXt, and normalization stabilizes training with pretrained weights.  
\end{itemize}

This strategy allows the model to learn features that are robust to rotations, scale changes, staining variations, and imaging artifacts, 
thereby improving generalization to unseen tumor domains and scanner conditions.

Validation transformations applied only center crop, resizing, and normalization to evaluate performance on realistic, unaltered data.

\subsection*{Training Protocol}
All experiments were conducted using a 5-fold cross-validation scheme. Each fold was trained with the following configuration:  

\begin{itemize} 
    \item \textbf{Loss function:} Binary Cross-Entropy with Logits.  
    \item \textbf{Optimizer:} AdamW with learning rate $1 \times 10^{-4}$.  
    \item \textbf{Learning rate scheduler:} Cosine Annealing LR with $T_{\text{max}}$ equal to the number of epochs and $\eta_{\min} = 10^{-7}$.  
    \item \textbf{Batch size:} 128.  
    \item \textbf{Number of epochs:} 20 per fold.  
    \item \textbf{Regularization:} Implicit via extensive histopathology-specific data augmentations and balanced sampling; early stopping was not explicitly used, but the best model per fold was saved based on validation Balanced Accuracy.  
    \item \textbf{Balanced sampling:} WeightedRandomSampler based on inverse class frequency to mitigate class imbalance.  
\end{itemize}

\subsection*{Evaluation Protocol}
Model performance was assessed using a 5-fold stratified cross-validation strategy. For each fold, the following evaluation setup was applied:

\begin{itemize}
    \item \textbf{Metric:} Balanced Accuracy, computed on the validation set of each fold.  
    \item \textbf{Model selection:} The best model per fold was saved based on the highest validation Balanced Accuracy.  
\end{itemize}

\subsection*{Reproducibility}
To ensure full reproducibility of our results, we applied the following measures:

\begin{itemize}
    \item \textbf{Random seed:} Fixed at 42 for all cross-validation splits.  
    \item \textbf{Explicit settings:} All preprocessing, data augmentations, batch size, learning rate, optimizer, and scheduler were fully specified.    
    \item \textbf{Model checkpoints:} Best model weights were saved per fold to allow replication of results.
\end{itemize}

\section*{Results}
ConvNeXt trained with histopathology-specific augmentations achieved strong and consistent performance across cross-validation folds. Using our methodology, the model ranked among the \textbf{top entries} in Track 2 of the MIDOG 2025 Challenge, achieving a \textbf{Balanced Accuracy of 0.8961} and \textbf{ROC AUC of 0.9561}. Across individual tumor domains, Balanced Accuracy ranged from 0.8843 to 0.9444, and ROC AUC from 0.9347 to 1.0000, demonstrating robust generalization across diverse tumor types.

\section*{Discussion}
Our findings show that ConvNeXt, a hybrid architecture combining convolutional layers with transformer-style design, is highly effective for mitotic figure classification. By randomly introducing diverse geometric, color, and noise-based augmentations, the model experiences a wide variety of tissue and scanner variations, which enhances generalization to unseen tumor domains. These results underscore the importance of domain-aware augmentation strategies in computational pathology.\\

Importantly, compared to foundation model approaches, whether frozen or fine-tuned, our lightweight ConvNeXt-based model demonstrates competitive performance while maintaining a significantly smaller computational footprint. This efficiency is particularly important in cellular-scale applications, where millions of cells per slide must be processed. By combining a compact architecture with aggressive, histopathology-specific augmentations and balanced sampling, the model achieves robust generalization across diverse tumor domains, offering a practical solution for high-throughput computational pathology without sacrificing accuracy.\\

Future work could explore additional domain adaptation techniques, including test-time augmentation \cite{shanmugam2021betteraggregationtesttimeaugmentation}, Macenko stain normalization \cite{inproceedings}, and further augmentation strategies. Systematically evaluating multiple adaptation approaches would help identify the most effective methods to improve generalization under domain shift, ensuring robustness across heterogeneous datasets.

\begin{acknowledgements}
We thank the MIDOG 2025 Challenge organizers for providing the dataset and evaluation platform, which support advances in automated mitotic figure analysis.

\end{acknowledgements}

\section*{Bibliography}
\bibliography{literature}

\end{document}